\newcommand{\appsection}[1]{\let\oldthesection\thesection
  \renewcommand{\thesection}{Appendix \oldthesection}
  \section{#1}\let\thesection\oldthesection}
 
\documentclass[12pt,fleqn,psfig]{article}
\usepackage{amsmath, amssymb, graphicx, float,appendix}

\begin{document}


\textwidth 6.3in \textheight 8.8 in \hoffset -0.5 in \voffset -0.3
in
\renewcommand{\theequation}{\thesection.\arabic{equation}}


\thispagestyle{empty}
\renewcommand{\thefootnote}{\fnsymbol{footnote}}

{\hfill \parbox{4cm}{
        Brown-HET-1620 \\
}}

\bigskip\bigskip\vskip6pt

\begin{center} \noindent
On exact statistics and classification of ergodic systems of integer dimension
\end{center}

\bigskip\bigskip\bigskip

\centerline{ \normalsize  Zachary Guralnik$^{1}$\footnote[1]{e-mail: zack@het.brown.edu}, Cengiz Pehlevan$^{2,1}$\footnote[2]{e-mail: cengiz@het.brown.edu} \footnote[4] {Current Addres: HHMI Janelia Farm Research Campus, 19700 Helix Dr., Ashburn, VA 20147},  Gerald Guralnik$^1$\footnote[1]{e-mail: gerry@het.brown.edu} }

\bigskip

\bigskip

\centerline{1. Department of Physics}
\centerline{Brown University}
\centerline{Providence, RI 02912}
\bigskip

\centerline{2. Harvard University}
\centerline{Center for Brain Science}
\centerline{Cambridge MA, 02138}
\bigskip


\bigskip\bigskip

\renewcommand{\thefootnote}{\arabic{footnote}}

\centerline{ \small Abstract}
\bigskip

\small

We describe classes of ergodic dynamical systems for which some statistical properties are known exactly.  These systems have integer dimension, are not globally dissipative, and are defined by a probability density  and a two-form.  This definition  generalizes the construction of  Hamiltonian systems by a Hamiltonian and a symplectic form.   Some low dimensional examples are given, as well as a discretized field theory with a large number of degrees of freedom and a local nearest neighbor interaction. We also evaluate unequal-time correlations of these systems without direct numerical simulation,  by Pad\'e approximants of a short-time expansion.  We briefly speculate on the possibility of constructing chaotic dynamical systems with non-integer dimension and exactly known statistics.  In this case there is no probability density, suggesting an alternative construction in terms of a Hopf characteristic function and a two-form.  

\newpage

\section*{Lead Paragraph}
{\bf Chaos is ubiquitous in nature. Studies of chaotic systems, however, are limited by the very few analytical tools available.  As evidenced by other areas of science, having ``toy models" with known exact results may provide deep insight into the nature of chaotic systems.  Here we construct ergodic dynamical systems with integer dimension for which exact statistics are known.  Chaos is frequent in such systems. Our method is a generalization of the construction of Hamiltonian flows from a Hamiltonian and a symplectic form. We define ergodic dynamical systems by a probability density and a two-form. This definition allows us to provide a classification of ergodic dynamical systems of integer dimension.
}

\section{Introduction}

A fundamental feature of chaos is the practical impossibility of predicting the state of a chaotic system arbitrarily far in the future. Beyond a certain time, statistical properties become of far more interest than the detailed evolution.  Long time numerical solutions are still of use to compute statistical properties. Obtaining statistical properties by direct numerical simulation has the disadvantage of being essentially an experimental approach.  As such, it yields no generalizable insight.  Unfortunately there are very few analytical tools available to analyze chaotic systems.  In particular, there are very few ``toy models" where exact results are known.  

 We will provide a means to construct an infinite class of ergodic systems for which the exact statistics are known.  Whether these are in fact chaotic rather than quasiperiodic must be checked, for instance by computation of Lyapunov exponents, however chaos is a commonly seen feature in ergodic systems. Our approach employs an inverse method introduced in \cite{G}, in which dynamical systems are constructed from a scalar function $\rho(\vec X)$ and a two form $B=B_{ij}dX^i\wedge dX^j$, defined on the phase space parameterized by $\vec X$.  This bears some resemblance to the construction of Hamiltonian flows from a Hamiltonian and a symplectic form.  Indeed, Hamiltonian flows arise as a special case. The scalar function in the inverse approach corresponds to a probability density, modulo normalization, the existence of which implies that the the toy models we obtain are never globally dissipative and have integer dimension.  While $\rho$ and $B$ are sufficient to define dynamical systems, ergodicity is generic but not guaranteed.  When ergodic,  the probability density is given by $\rho$, up to a normalization, within some region of support which also depends  on $B$.  The dependence of the domain of support on $B$ is often topological in character, being invariant under continuous deformations of $B$.
In addition to some examples of exactly soluble models with a small number of degrees of freedom, we also give an example with a large number of degrees of freedom and nearest neighbor interactions, akin to a discrete version of a non-linear field theory.

Many chaotic systems of interest are dissipative with a non-integral dimension, so that the invariant measure on phase space can not be written in terms of a probability density function, $d\mu(\vec X) \ne \rho(\vec X) d^N(\vec X)$.  We speculate that it may also be possible to reverse engineer such systems, starting with a two-form and a Hopf-characteristic function $Z(\vec J)$ instead of a probability density.  The Hopf function is the Fourier-Stieltjes transform of the invariant measure on phase space \cite{Hopf,Frisch}.  While  the invariant distribution may be a very complex fractal-like object, the Hopf function is generally analytic except at infinity, with asymptotic properties determined by the fractional dimension \cite{GCG} (see also \cite{PerSjolin,Erdogan,Edgar,KPG} for related work).

While the invariant
distribution of an ergodic dynamical system yields equal time correlations,  it contains little information about temporal correlations.  Indeed there exist many dynamical systems with the same invariant distribution,  or invariant distributions which are equivalent up to normalization within some domain,  but which have very different temporal correlations.  For the systems we consider,  
information about temporal correlations is a function of both the invariant measure (or $\rho$) and the two form  $B$.
We will describe initial efforts to compute temporal correlations for chaotic systems which have been reverse engineered from a probability distribution and a two form,  without recourse to any time series simulation.  
 We will have a modicum of success computing temporal auto-correlations $<X^i(0)X^j(\tau)>$ using Pad\'e approximants to the  expansion in $\tau$, with coefficients defined in terms of $\rho$ and $B$.

\section{Classification}

The trajectory of an ergodic dynamical system  may be characterized by an invariant measure on phase space $d\mu(\vec X)$ such that time averages over the trajectory are equivalent to phase space averages:
\begin{align}
\lim_{T\rightarrow\infty}\frac{1}{T}\int_0^T \,dt\, f(\vec X(t)) = \int d\mu(\vec X) f(\vec X).
\end{align}
It is often the case, as in dissipative chaotic systems, that the measure has fractal-like properties such as fractional Hausdorff dimension, and can not be expressed in terms of continuous differentiable functions.  
However in other cases which are not globally dissipative the measure has integer dimension and can be written as 
\begin{align}
d\mu(\vec X) = \rho(\vec X) d^N\vec X.
\end{align}
One may think of $\rho$ as a probability density.  Probability conservation implies that  
\begin{align}
\vec\nabla\cdot(\rho\vec v(\vec X)) = 0,
\end{align}
where $v$ is the velocity vector which defines the dynamical system:
\begin{align}\label{3}
\frac{d\vec X}{dt} = \vec v(\vec X).
\end{align}

It will be convenient to express \eqref{3} in terms of differential forms,
\begin{align}
d^*(\rho v) = 0,
\end{align}
where $v$ is a one-form, $d$ the exterior derivative and ${}^*$ the Hodge star.
Thus locally 
\begin{align}
{}^*(\rho v) = d{\cal A},
\end{align}
where for $N$ degrees of freedom, ${\cal A}$ is an $N-2$ form.  For systems of large $N$, it is often  more convenient to work with the two-form $B={}^*{\cal A}$.  

In \cite{G} it was pointed that these observations can be used to reverse engineer dynamical systems. Here, we will reverse engineer ergodic dynamical  systems with integer Haussdorf dimension and polynomial governing equations  by 
choosing $\rho$ and ${\cal A}$ such that $v = {}^*d{\cal A}/\rho$ is polynomial.
There are a vast number of ways to do this, of which we enumerate a few below.   It must be emphasized  that not all choices of $\rho$ and ${\cal A}$ lead to a chaotic dynamical system, nor will $\rho$ always correspond to an ergodic distribution.  In general, $\rho$ will be an invariant distribution, meaning that initial conditions randomly distributed with probability density $\rho$ remain so under time evolution.   This does not necessarily mean that time evolution starting from a single initial condition yields a trajectory visiting regions of phase space with frequency given by $\rho$.   Frequently, the invariant measure of a  trajectory will be  $d\mu(\vec X)= \tilde \rho(\vec X)d^N\vec X$  where $\tilde \rho$  is proportional to $\rho$ within some domain and vanishes elsewhere.  

The characterization of dynamical systems by a two-form and a scalar function of the phase space may remind the reader of symplectic dynamics.  Indeed, 
one can obtain Hamiltonian flows as a special case, with $\rho$ a function of the Hamiltonian and ${}^*\Omega$ proportional to the symplectic form:
\begin{align}
\rho &= H(q_i,p_i)\\
{}^*\Omega &=\frac{1}{2} \omega,
\end{align}
where $\omega$ is the symplectic form,
\begin{align}
\omega=\sum_i dq_i\wedge dp_i.
\end{align}
If the dynamics is ergodic, $\rho$ is the probability distribution up to a a normalization factor within the domain of the invariant set.  For Hamiltonian dynamics,  the flows are restricted to constant $H$ and the probability distribution is a constant.

\subsection{Polynomial class}\label{PPL}

The simplest class of $N-2$ forms ${\cal A}$ and distributions $\rho$ leading to polynomial $v$ is obtained by  
choosing  polynomial $\rho$ and 
\begin{align}
{\cal A} = \rho^2\Omega,
\end{align} 
where $\Omega$ is also polynomial.  Then
\begin{align}\label{polyclass}
{}^*v=\rho d\Omega+ 2d\rho\wedge\Omega.
\end{align}

As an example consider the dynamical system defined by
\begin{align}
\rho &= (1-x^4-y^2-z^6) \\
\Omega &= (xyz)dy + (y^2)dz ,
\end{align}
giving rise to the dynamical system:
\begin{align}\label{purepol}
v_x&=13 z^6xy-6y^3-xy+2y+yx^5-2yx^4-2yz^6+y^3z\nonumber\\
v_y&=8x^3y^2\nonumber\\
v_z&=-9x^4yz +yz -yz^7 -y^3z .
\end{align}
One can check numerically that this leads to ergodic dynamics on the domain
\begin{align}\label{domain}
x^4 + y^2 +z^6 <1, \qquad y>0,\qquad z&>0,
\end{align}
with statistics characterized by $\rho$, suitably normalized. 
Various equal time correlations are shown in Table \ref{T1}, computed both by a time series simulation and using the proposed exact invariant measure, showing precise agreement.  As we will see later, the auto-correlation falls off with time (and the power spectral density is broad-band) ruling out quasiperiodicity.
The top Lyapunov exponent is found numerically to be  positive ($\lambda\approx 0.2$) so that the dynamics is indeed chaotic.  

Note that while $\rho$ takes a polynomial form within the domain of support, there is no requirement for $\rho$ to be analytic in the entire phase space.  Clearly the domain of support always lies within the region $\rho>0$, however there are frequently  additional constraints which depend on the two-form $B$ (or the velocity $v$) as well as $\rho$.

\begin{table}[H]
\centerline{
\begin{tabular}{ l  c  c}
Moments & Dynamics & Monte Carlo \\
\hline $<y>$ & $0.3449$ & $0.3451$ \\
 $<z>$ & $0.4223$ & $0.4136$ \\  $<x^2>$ & $0.2066$ &
$0.2068$ \\  $<y^2>$ & $0.1720$ & $0.1717$ \\
 $<z^2>$ & $0.2387$ & $0.2324$ \\  $<yz>$  &
$0.1403$ & $0.1386$ \\
 $<y^3>$  & $0.1014$ & $0.1011$ \\
 $<z^3>$ & $0.1531$ & $0.1493$ \\
 $<x^2y>$ & $0.06562$ & $0.06543$ \\
 $<y^2z>$ & $0.06836$ & $0.06721$ \\
 $<yz^2>$ & $0.07670$ & $0.07571$ \\
 $<x^4>$  & $0.08531$ & $0.08579$ \\
 $<x^2yz>$ & $0.02607$ & $0.02564$\\
 $<x^2y^2>$ &$0.03056$ & $0.03021$\\ 
$<x^6y^4z^2>$ & $0.0002604$ & $0.0002482$ 
\end{tabular}
} 
\caption{\label{T1}The entries in the column labeled `dynamics' were
obtained by numerical simulation of the dynamical system
\eqref{dyns} over a time duration $100000$ , sampling at intervals
$\Delta t=0.01$, starting from the initial conditions
$(x,y,z)=(0.5,0.5,0.5)$. The column labeled `Monte Carlo' was
obtained by a hit and miss Monte Carlo calculation using the
proposed exact distribution with 400000 accepted data points.}
\end{table}




\subsection{Polynomial--Exponential class}

Let us now consider probability distributions of the form 
\begin{align}\label{expdist}
\rho = Pe^{-Q} ,
\end{align}
where $P$ and $Q$ are polynomial. To insure that
$v$ is polynomial, one chooses an $N-2$ form
\begin{align}
\Omega = \xi e^{+Q} ,
\end{align}
where $\xi$ is a polynomial $N-2$ form.  Then
\begin{align} \label{expvel}{}^*v=
Pd\xi+2dP\wedge\xi-PdQ\wedge\xi .
\end{align} 

As an example with $N=3$, consider
\begin{align}
P &= (1-x^2-z^6)\, , \nonumber\\
Q &= x^4+y^4+z^2+3xyz\, , \label{nonp} \\
\xi &= xyzdy+y^2dz\, .\nonumber
\end{align}
It is not hard to see that initial conditions with $y>0$ and $z>0$ remain within this domain. 
The probability distribution is in fact given by $\rho$ within this domain, up to a  normalization, and zero elsewhere.
Comparing connected moments\footnote{$<xy>_c\equiv <xy>-<x><y>$,\\
\indent\hskip8pt $<xyz>_c \equiv
<xyz>-<xy>_c<z>-<xz>_c<y>-<yz>_c<x>-<x><y><z>$, etc...} associated
with the probability distribution
\begin{align}\label{expdist1}
\rho &= (1-x^2-z^6)e^{-(x^4+y^4+z^2+3xyz)}\,\,\,{\rm for}\,\,\,
x^2+z^6\le
1,\,\,\, y>0,\,\,\,z>0\nonumber\\
\rho &= 0\,\,\,{\rm elsewhere}.
\end{align} with moments
generated by the chaotic trajectory suggests that the two agree.
(see Table \ref{expTable}).

\begin{table}[H]
\centerline{
\begin{tabular}{ l c  c}
Moments & Dynamics & Monte Carlo \\ \hline $<x>$ & $-0.08497$
& $-0.09152$ \\$<y>$ & $0.4898$ & $0.5046$
\\  $<z>$ & $0.3712$& $0.3813$ \\ 
$<x^2>_c$ & $0.1697$ & $0.1685$ \\  $<y^2>_c$ & $0.1052$ & $0.1028$ \\
 $<z^2>_c$ & $0.06145$ & $0.05809$ \\
 $<xy>_c$ & $-0.01830$ & $-0.01887$ \\
 $<xz>_c$ & $-0.01304$ & $-0.01221$ \\
 $<yz>_c$ & $0.001292$ & $0.0033309$ \\
 $<x^3>_c/(<x^2>_c)^{3/2}$ & $0.1953$ & $0.1946$ \\
 $<y^3>_c/(<y^2>_c)^{3/2}$ & $0.4361$ & $0.4119$ \\
 $<z^3>_c/(<z^2>_c)^{3/2}$ & $0.3828$ & $0.3027$ \\
 $<xyz>_c/\sqrt{<x^2>_c<y^2>_c<z^2>_c}$ & $-0.07400$ & $-0.07370$ \\
 $<x^4>_c/<x^2>_c^2$ & $-0.7019$ & $-0.7433$ \\
\end{tabular}
}
\caption{\label{expTable}Entries in the column labeled `dynamics' were obtained by
numerical simulation of the dynamical system over a time duration
$40000$, sampling at intervals $\Delta t= 0.1$, starting from the
initial conditions $(x,y,z)=(0.1,1.1,0.4)$. The column labeled
`Monte Carlo' was obtained by a hit and miss Monte Carlo calculation
using the probability distribution \eqref{expdist1} with 400000
accepted data points.}
\end{table}

\subsection{Other classes}

The inverse
approach generalizes to  probability distributions with significantly
more complicated analytic structure.  To illustrate, consider a distribution of the form
\begin{align}\label{complx}
\rho = \frac{P_1}{P_2}e^{-P_3/P_4}\, ,
\end{align}
with polynomial $P_i$, together with the $N-2$ form
\begin{align}
{\cal A} = P_1^2P_4^2e^{-P_3/P_4}\xi\, ,
\end{align}
where $\xi$ is a polynomial $N-2$ form.  The resulting velocity field
$v={}^*d{\cal A}/\rho$ is polynomial. One must check that the dynamics is
ergodic when restricted to some non-trivial
domain of support.  Although \eqref{complx}
may have real poles and essential singularities,  these are not pathological  if they are integrable or lie outside the domain of support of the ergodic distribution.

\subsection{Lattice fields with local interactions}

In this section we describe a generalization of dynamical systems in the polynomial class described in section \ref{PPL} to discretized field theories with nearest neighbor interaction.  We shall first solve the simpler problem of finding the two form and probability distribution which describe $N$ un-coupled copies of a dynamical system described by \eqref{polyclass} with $p$ degrees of freedom, such that the total number of degrees of freedom is $pN$. We shall label the degrees of freedom $X_{iI}$ where $i=1\cdots N$ and $I=1\cdots p$.   The dynamics is given by 
\begin{align}\label{ultralocal}
{}^*v=\rho d\Omega+ 2d\rho\wedge\Omega.
\end{align}
with probability density
\begin{align}
\rho=\prod_{i=1}^N \rho_i(X_{i1} \cdots X_{ip})
\end{align}
and $Np-2$ form
\begin{align}
\Omega = \frac{1}{\prod_{j\ne 1} \rho_j}&\Omega_1 \wedge \omega_2\wedge \omega_3\wedge\cdots \wedge \omega_N \nonumber\\  
+ \frac{1}{\prod_{j\ne 2}\rho_j} &\omega_1 \wedge \Omega_2 \wedge \omega_3 \wedge \cdots \wedge \omega_N \nonumber\\
+&\cdots \nonumber\\
+ \frac{1}{\prod_{j\ne N}\rho_j} &\omega_1  \wedge \cdots \wedge\omega_{N-1} \wedge\Omega_N
\end{align}
Here $\omega_i$ are the volume forms associated with each lattice site, 
\begin{align}
\omega_i= dX_{i1}\wedge dX_{i2} \cdots\wedge dX_{ip}\, .
\end{align}
while each $\Omega_i$ is a $p-2$ form associated with original system in the polynomial class, which has been copied at each lattice site.  It is not hard to verify that \eqref{ultralocal}
gives $N$ un-coupled copies of dynamical system defined by
\begin{align}\label{localv}
{}^{*p}v_i=\rho_i d\Omega_i+ 2d\rho_i\wedge\Omega_i,
\end{align}
where ${}^{*p}$ indicates the hodge dual with respect to the subspace spanned by $X_{i1}\cdots X_{ip}$.
To make things concrete,  we may consider $N$ copies of the dynamical system \eqref{purepol}.
In this case,
\begin{align}\label{UL}
\rho_i &= 1-x_i^4-y_i^2-z_i^6\nonumber\\
\Omega_i &= (x_iy_iz_i)dy_i+(y_i^2)dz_i \nonumber\\
\omega_i &= dx_i \wedge dy_i \wedge dz_i
\end{align}
Here $p=3$, and $X_{i1}\equiv x_i$,  $X_{i2}\equiv y_i$ and $X_{i3}\equiv z_i$.

Next one can modify the probability density $\rho$ and the $Np-2$ form $\Omega$ such 
that neighboring degrees of freedom interact, while keeping $v$ polynomial.
To this end, we take 
\begin{align}\label{coupledrho}
\rho = \prod_i \rho_i \prod_{jk}\rho_{jk},
\end{align}
where $\rho_{jk}$ is a function which couples degrees of freedom at the sites $j,k$, meaning it cannot be written as a product of a function of degrees of freedom at $j$  and another function of degrees of freedom at site $k$.
We then choose
\begin{align}
\Omega = \frac{1}{\prod_{i\ne 1} \rho_i\prod_{j\ne1,k\ne 1} \rho_{jk}}&\Omega_1 \wedge \omega_2\wedge \omega_3\wedge\cdots \wedge \omega_N \nonumber\\  
+ \frac{1}{\prod_{i\ne 2}\rho_i \prod_{j\ne 2,k\ne 2} \rho_{jk}} &\omega_1 \wedge \Omega_2 \wedge \omega_3 \wedge \cdots \wedge \omega_N \nonumber\\
+&\cdots \nonumber\\
+ \frac{1}{\prod_{i\ne N}\rho_j\prod_{j\ne N,k\ne N} \rho_{jk}} &\omega_1  \wedge \cdots \wedge\omega_{N-1} \wedge\Omega_N
\end{align}

The dynamics obtained from \eqref{ultralocal} is then 
\begin{align}
{}^{*p}v_i = \left(\prod_j \rho_{ij}\right){}^{*p}v_i^{(0)} + 2\rho_id_i\left(\prod_j \rho_{ij}\right)\wedge \Omega_i
\end{align}
where $d_i$ is the exterior derivative with respect to degrees of freedom at the lattice site $i$ only, and $v_i^{(0)}$ is the velocity in the absence of coupling between sites, given by \eqref{localv}.

As a specific example, we again consider $\Omega_i$ and $\rho_i$ given by \eqref{UL}, with
\begin{align}\label{crossrho}
\rho_{ij} = 1-\gamma(y_i -y_j)^2
\end{align}
with $j=i\pm 1$, and cyclic symmetry $i+N \equiv i$, such that the system lives on a discretized circular lattice:
\begin{align}
\prod_{ij} \rho_{ij} = (1-\gamma(y_1 -y_2)^2)&(1-\gamma(y_2 -y_3)^2)\cdots \nonumber\\ 
\cdots&(1-\gamma(y_{N-1} -y_N)^2)(1-\gamma(y_N -y_1)^2)
\end{align}
Then the  equations of motion are 
\begin{align}
v_{x,i} = &\rho_{i-1,i}\rho_{i,i+1}  v_{x,i}^{(0)} + 2\rho_i y_i^2 \frac{\partial}{\partial y_i} \left(\rho_{i-1,i}\rho_{i,i+1}\right) \nonumber\\
v_{y,i} =  &\rho_{i-1,i}\rho_{i,i+1}  v_{y,i}^{(0)} \nonumber\\
v_{z,i} = &\rho_{i-1,i}\rho_{i,i+1}  v_{z,i}^{(0)}
\end{align}
with $\rho_{ij}$ given in \eqref{crossrho}, $\rho_i$ given in \eqref{UL} and  $v^{(0)}$ the velocities of the non-interacting case, as in \eqref{purepol} copied at each lattice site.

Comparing correlation functions obtained from the probability distribution \eqref{coupledrho} (suitably normalized), and from the dynamics shows close agreement for certain, but not all, values of $N$ and $\gamma$.  Of particular interest are the connected correlations between different lattice sites, such as 
\begin{align}
<y_iy_{i+1}>_c \equiv <y_i y_{i+1}> - <y_i><y_{i+1}>.
\end{align}
In light of the discrete translational symmetry, $i\rightarrow i+1$, we have averaged over the index $i$ in evaluating such expressions.  Note that it is possible that the ergodic domain violates this symmetry, even if the equations of motion and the probability distribution function do not.  However, we have yet to stumble on such behavior.  

Results for $N=4$ and two different vallues of $\gamma$ are shown in Table \ref{SmallTable}. 

\begin{table}[H]
\centerline{
\begin{tabular}{l c c c c}
Moments   &  Dynamics  &  Monte Carlo &   Dynamics  &   Monte Carlo  \\ 
& $\gamma=1$&  $\gamma=1$& $\gamma=10$& $\gamma=10$\\
\hline
$<y_i>$ & $0.3344$ & $0.3324$ & $0.2739$ & $0.2721$
\\  $<z_i>$ & $0.4155$& $0.4146$ & $0.4159$ & $0.4203$\\ 
$<x_i^2>_c$ & $0.2079$ & $0.2091$ & $0.2166$ & $0.2166$ \\
$<z_i^2>_c$ & $0.2322$ & $0.2337$ & $0.2362$ & $0.2398$\\
 $<y_i^2>_c$ & $0.1632$ & $0.1553$ & $0.1015$ & $0.0987$\\
 $<y_iy_{i+1}>_c$ & $0.005089$ & $0.005224$ & $0.01729$ & $0.01702$\\
 $<y_iy_{i+2}>_c$ & $0.001112$ & $0.001182$ & $0.01529$ & $0.01484$\\
\end{tabular}
}
\caption{\label{SmallTable}Entries in the column labeled `dynamics' were obtained by
numerical simulation of the dynamical system over a time duration
$200000$, sampling at intervals $\Delta t = 0.02$, starting from random 
initial conditions. The column labeled
`Monte Carlo' was obtained by a hit and miss Monte Carlo simulation
using the probability distribution \eqref{coupledrho} with $5235915$
accepted points.}
\end{table}

We have also obtained good agreement for much larger $N$.
Results for $\gamma=0.5$ and $N=32$ are shown in Table \ref{BigTable}. 

\begin{table}[H]
\centerline{
\begin{tabular}{ l c  c}
Moments & Dynamics & Monte Carlo \\ \hline
 $<y_i>$ & $0.3381$ & $0.3392$
\\  $<z_i>$ & $0.4130$& $0.4135$ \\ 
$<x_i^2>_c$ & $0.2079$ & $0.2075$  \\
 $<z_i^2>_c$ & $0.2322$ & $0.2325$ \\
 $<y_i^2>_c$ & $0.1632$ & $0.1639$ \\
 $<y_iy_{i+1}>_c$ & $0.002708$ & $0.002681$ \\
\end{tabular}
}
\caption{\label{BigTable}Entries in the column labeled `dynamics' were obtained by
numerical simulation of the dynamical system over a time duration
$200000$, sampling at intervals $\Delta t = 0.02$, starting from random 
initial conditions. The column labeled
`Monte Carlo' was obtained by a Metropolis Monte Carlo simulation
using the probability distribution \eqref{coupledrho} with $1000000$
lattice sweeps.}
\end{table}
In performing the Monte Carlo evaluation, we have assumed that the ergodic domain \eqref{domain}, which contains our initial conditions, continues to hold for the degrees of freedom at each lattice site.  For sufficiently large of values of $\gamma$,  we found that the Monte Carlo and dynamical results differ,  presumably because the domain \eqref{domain} is no longer valid, or the system ceases to be chaotic.

By considering a suitable large $N$ limit, it should be possible to reverse engineer continuous field theories exhibiting chaotic behavior for which the statistics are known exactly.  Some such theories may resemble turbulent fluids in some respects. Of course these will be of a somewhat special type, having integer rather than fractal dimension due to the existence of a probability density function, and exhibiting dissipation in some regions of phase space but not in others.

\section{Temporal correlations}

Thus far we have described an analytic inverse approach to construct ergodic dynamical systems from a probability distribution and a two form. By itself, the probability distribution contains no information about temporal correlations.  We will attempt below to extend this analytic approach to the computation of un-equal time correlations, which are dependent also on the two-form.

 Equal time moments are determined by the invariant
distribution, subject to constraints on the domain of support
required for ergodicity. The dependence of these constraints on the
two-form ${}^*\Omega$ is often topological in character, as smooth variations of
the two-form do not necessarily change the constraints.  On the
other hand, un-equal time correlation depend directly on the
two-form as well as the invariant distribution, through the
dependence of $\vec v(\vec x)$ on these quantities.
Indeed,  there exist different chaotic dynamical systems with the same invariant measure, either globally or  in certain domains of phase space.  Information about temporal correlations requires both the scalar function $\rho$ (or Hopf function $Z$ if $\rho$ does not exist)  and the two-form ${}^*\Omega$, which together define the dynamics.
Below we describe initial efforts to compute temporal correlations for chaotic systems which have been reverse engineered from a probability distribution and a two form,  without any time series simulation.  

Consider the auto-correlation function of the phase space variable $x$,
\begin{align} G(\Delta)\equiv<x(t)x(t+\Delta)> - <x(t)>^2 .\end{align}
where brackets indicates time averaging. Time averages of functions of phase space over an ergodic chaotic
trajectory are equal to averages with respect to an invariant
measure over phase space:
\begin{align} 
< f(\vec x(t))> = \int d\mu(\vec X) f(\vec X).
\end{align} 
Noting that $\vec x(t+\Delta)$ is a function of $\vec
x(t)$, the auto-correlation of a phase space variable $x_i$ can be
written as
\begin{align}\label{theqn}
G_i(\Delta)=\int d\mu(\vec X) f_\Delta(\vec X) - \left(\int
d\mu(\vec X) X_i\right)^2 ,
\end{align}
where 
\begin{align}\label{fps} f_\Delta(\vec X) \equiv
X_ix_i(\Delta)\, , \,\,\,\,\, {\rm with}\,\, \vec x(0)= \vec X.
\end{align} 
While chaos precludes numerical prediction of $\vec x(\Delta)$ for sufficiently large $\Delta$, there is no fundamental obstruction to computing the quantity  $\int d\mu(\vec X) X_i x_i(\Delta)$ for all  $\Delta$. 
 In a conventional approach, the auto-correlation is evaluated
by a long duration numerical simulation of the equations of motion,
taking the $t$ average of $x(t)x(t+\Delta)$, or by shorter time averages in which one also sums over many randomly chosen initial conditions.  We view this as a brute force approach, with no possibility of obtaining information about the analytic structure of $G(\Delta)$.
 
We will attempt to compute the autocorrelation by Pad\'e approximants derived from the Taylor-Mclaurin series for $G_i(\Delta)$.  The Taylor-Mclaurin series has a non-zero radius of convergence, due to the assumed absence of singularities of $\vec x(\Delta)$ on the real $\Delta$ axis. To obtain the series expansion for $G_i(\Delta)$, let us first expand $f_\Delta(\vec X)$  in $\Delta$ , assuming that $x_i$ has been shifted
such that $<x_i(t)>=0$:
\begin{align}
F_{\Delta}(\vec X) \sim x_i(0)\left(x_i(0)+\Delta \dot x_i(0)
+\frac{1}{2} \Delta^2 \ddot x_i(0)\cdots \right)\, , \,\,\,{\rm
with}\,\,\, \vec x(0)=\vec X.
\end{align}
Note that $x_i\frac{d^nx_i}{dt^n}$ can be written as a total time
derivative for odd $n$, \begin{align}x_i \frac{dx_i}{dt}  =
\frac{d}{dt}\left(\frac{1}{2}x_i^2\right),\,\,\,
x_i\frac{d^3x_i}{dt^3} =
\frac{d}{dt}\left(x_i\frac{d^2x_i}{dt^2}-\frac{1}{2}\left(\frac{dx_i}{dt}\right)^2\right)\,,\,\,{\rm
etc}\, ,\end{align} while for even $n=2m$,
\begin{align} x_i\frac{d^nx_i}{dt^n} =
\frac{d}{dt}\left(g_n\right) +(-1)^{m}\left(\frac{d^m
x_i}{dt^m}\right)^2\, ,
\end{align}
where we will not bother to specify $g_n$, since total time
derivatives vanish when averaged over the 
trajectory of a bounded system;
$<\frac{d}{dt}g_n(\vec x(t))>=0$. Thus
\begin{align}\label{sers}
G_i(\Delta)&=\int d\mu(\vec X)F_\Delta(\vec X) \sim \sum_{m=0}^\infty
\alpha_m\Delta^{2m},\,\,\,\\{\rm with}\,\,\, \alpha_m &\equiv
(-1)^m\frac{1}{2m!}\left<\left(\frac{d^mx_i(t)}{dt^m}\right)^2\right>\,
.\end{align} Using the equations of motion $\frac{dx_i}{dt} =
v_i(\vec x)$, the terms $\left(\frac{d}{dt}\right)^m x_i$ can be
related to functions on phase space. For example
\begin{align}\frac{d^2x_i}{dt^2}= \frac{dv_i}{dt}
 = \sum_j v_j(\vec x)\frac{\partial v_i(\vec x)}{\partial x_j}\, ,\end{align} so that \begin{align}\alpha_2
 = \left<\left(\frac{d^2x_i}{dt^2}\right)^2\right> = \int d\mu(\vec
 X)\left(\sum_j v_j(\vec X)\frac{\partial v_i(\vec X)}{\partial
 X_j}\right)^2\, .\end{align} In the subsequent section, we evaluate the coefficients $\alpha_m$ for a specific model for which the invariant measure $d\mu(\vec X)$ is exactly known,
followed by a Pad\'{e} resummation of \eqref{sers}.

\subsection{An example}

We revisit the chaotic dynamical system described in section \ref{PPL}
with $\rho$ and
$\Omega$ given by
\begin{align}\label{deft} \rho &=1-x^4-y^2-z^6\, , \nonumber \\
\Omega &= (xyz)dy + (y^2) dz\, ,
\end{align}
yielding
\begin{align}\label{dyns}
&v_x=13z^6xy-6y^3-xy+2y+yx^5-2yx^4-2yz^6+y^3x\, , \nonumber \\
&v_y=8x^3y^2\, , \\
&v_z=-9x^4yz+yz-yz^7-y^3z \, . \nonumber \end{align} Initial
conditions with $\rho(x,y,z)>0$, $y>0$ and $z>0$ give rise to
chaotic trajectories whose statistics are described by an invariant
distribution $\tilde \rho = \rho$ (up to normalization) within the
domain $\rho(x,y,z)>0,\, y>0,\, z>0$ and $\tilde \rho =0$ elsewhere.

\begin{figure}
\includegraphics{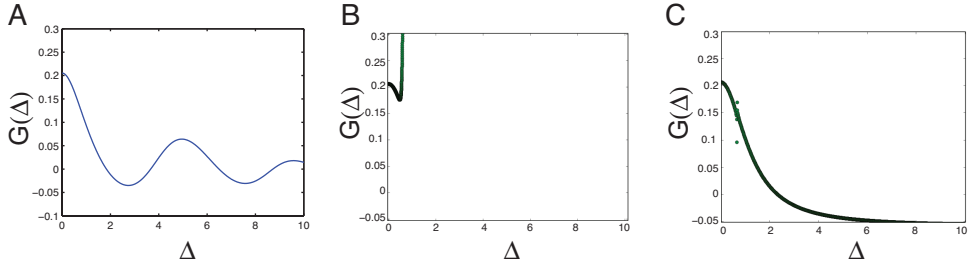}
\caption{\label{Fig1} Calculation of $G(\Delta)$ A) by the
numerical time series simulation, B) by 9'th order
Taylor-Maclaurin series  and C) by $(4|4)$ Pad\'{e}
approximant.  The Pad\'{e} result is a smooth
continuous curve, with the exception of a very small neighborhood of
the point $\Delta =0.6529$, at which there is a spurious pole.
We speculate that improved knowledge of the large $m$ behaviour of the series coefficients $\alpha_m$ could allow for more effective summation methods than a simple Pad\'{e} approximant. }
\end{figure}

Consider the auto-correlation $G_x(\Delta)$ for the variable $x$ on
these  trajectories.  Using \eqref{theqn};
\begin{align}\label{example}
G_x(\Delta) = \int dX dY dZ \, \tilde\rho(X,Y,Z)\,
f_\Delta(X,Y,Z) - \left(\int dX dY dZ \, \tilde\rho(X,Y,Z)\,
X\right)^2\, ,
\end{align}
where
\begin{align}
F_\Delta(X,Y,Z) \equiv Xx(\Delta)\, ,\end{align} with
\begin{align} x(0)=X,\,\,y(0)=Y,\,\,z(0)=Z\, .
\end{align}
We have calculated the Taylor-Mclaurin series coefficients
$\alpha_0\cdots\alpha_8$ of $G_x(\Delta)$ using computer algebra and Monte-Carlo simulation 
to evaluate integrals with respect to the invariant measure 
$d\mu(\vec X)=\tilde\rho(\vec X)$.  The
 $(4|4)$ Pad\'{e} approximant to $G(\Delta)$ is a ratio of polynomials
of degree 4 whose first 9 Taylor-Maclaurin series coefficients are
$\alpha_0\cdots\alpha_8$. The $(4|4)$ Pad\'{e} approximant to
$G_x(\Delta)$ is plotted in Figure \ref{Fig1}, along with the 9'th order
Taylor-Maclaurin series result and the result of direct numerical
simulation by a long run of the dynamical system. Note that the 9-th
order Taylor-Maclaurin series begins to differ markedly from the
result of direct numerical simulation at $\Delta\sim 0.35$, well
before the first zero, whereas the $(4|4)$ Pad\'{e} approximant
gives accurate result for much larger values of $\Delta$, and is an
acceptable approximation up to a neighborhood of the first zero,
$\Delta\sim 2$. 

Higher order Pad\'{e} approximants, or more sophisticated resummation methods,  might provide a good approximation for larger
values of $\Delta$.  The initial results
are encouraging, suggesting that the auto-correlation may be
computed without any direct simulation of the dynamical system. This has the advantage of replacing an
`experimental' approach to calculating auto-correlation functions by direct simultation
with a purely theoretical approach based on exact integral
expressions for the Taylor-Mclaurin expansion. The large order behavior of
the Taylor-Mclaurin expansion, if one can determine it, could yield very interesting results about the analytic structure and asymptotics of the auto-correlation.

Chaotic power spectra are expected to have a non-zero exponentially
small component at high frequency \cite{FrischMorf,Sigeti}, $\tilde
G(\omega)\sim \exp(-\alpha\omega)$. The time-scale $\alpha$ is
determined by the proximity of the nearest singularity of the
auto-correlation $G(\Delta)$ to the real $\Delta$ axis. Note that
there can not be any singularities on the real $\Delta$ axis, as it
is assumed that the time evolution of the dynamical system does not
encounter any singularities.  Due to the tendency of chaotic systems
to `forget' their initial conditions, one expects singularities of
$G(\Delta)$ near the real axis to occur for small values of
$Re(\Delta)$.  This suggest that the high frequency behavior of the
power spectral density can be extracted from the small $\Delta$
behavior of the auto-correlation.  It may therefore be possible to
use a relatively low order Pad\'{e} approximant to the
auto-correlation to get an estimate of the parameter $\alpha$. Some
care must be taken, since the poles of the Pad\'{e} approximant do
not necessarily correspond to the true analytic structure.  In fact
the $(4|4)$ Pad\'{e} approximant we have computed here has two poles
which are likely both spurious, including one on the positive real
axis which must be spurious.  These poles are extremely close to
zeros of the Pad\'{e} approximant.  There is one pole which is not
near any zero, at $\Delta=1.24 i$, suggesting $\alpha\approx 1.24$
as a crude first approximation.
\\

\section{Systems with non-integer dimension}

While we have obtained ergodic (and generically chaotic) dynamical systems with exactly known statistical properties,  these  are non-dissipative, due to the existence of an everywhere finite probability distribution function.  A simple argument shows that such a function can not exist in the presence of dissipation. If an everywhere finite $\rho$ did exist, it would satisfy $\frac{d\rho}{dt} = -\rho \vec\nabla\cdot\vec v >0$ along any trajectory, where $\vec v(\vec X)$ is the velocity $\frac{d\vec X}{dt}$, which is inconsistent with Poincar\'e recurrence.  The systems we have constructed are dissipative  in some regions of the  orbit but anti-dissipative (satisfying
$\frac{d\rho}{dt} = -\rho \vec\nabla\cdot\vec v <0$) in others.  Yet many, if not most, chaotic orbits of physical interest are globally dissipative.  Thus it would be very interesting to have an  inverse method to obtain dissipative chaotic systems with exactly known statistics.

Dissipative chaotic orbits are charactarized by a measure $d\mu(\vec X)$ on phase space, which can not be written in terms a probability density, $d\mu(\vec X)\ne\rho(\vec X)d^n\vec X$,  and which may have remarkable geometric properties such as a non-integer dimension $d<n$.  While there are no simple expressions for such fractal-like measures, the Fourier-Stieltjes transform,
\begin{align}
Z(\vec J)\equiv\left< \exp\left(i{\vec J}\cdot{\vec X(t)}\right) \right>=\int d\mu(\vec X)\exp\left(i{\vec J}\cdot\vec X\right),
\end{align}
 is generally $C^{\infty}$, with derivatives at $\vec J=0$ corresponding to equal time correlation functions.  Thus it might be possible to reverse engineer dissipative chaotic dynamical systems with exactly known statistics by starting with a Hopf function $Z(\vec J)$ and a two form.  This is  analagous to the construction described above, except that the probability density is replaced with the Hopf function.  

The feasibility of such an inverse approach is unknown to us at present, and there will be a number of constraints that the Hopf function $Z(\vec J)$ must satisfy at the outset.  In particular $Z(\vec J)$ must be constructed so as to correspond to a non-integer dimension $d<n$. This requirement is a constraint on the large $|\vec J|$ asymptotics \cite{GCG}.  Note that the absence of probability density implies that the Fourier transform  $\int d^n\vec X Z(\vec J)\exp(i\vec J\cdot\vec X)$ can not converge. Indeed, it was suggested in \cite{GCG} that the Haussdorf dimension is, for many systems, given by  the maximum value  $s\le n$ such that the integral
 \begin{align}\label{defcon}I_s=\int_{|J|>\epsilon} d^n J\, |J|^{s-n} |Z(\vec J)|^2
\end{align}converges.

\section{Conclusions}

 While time series simulation of chaotic dynamical systems is the usual
method to compute their statistics, it is difficult to
obtain theoretical insight from such an approach. Furthermore,
direct time series simulation places extreme or prohibitive demands on
computational resources for systems with a very large number of
degrees of freedom. We have shown that it is possible to reverse engineer large classes 
of ergodic, and generally chaotic, dynamical systems with an arbitrary number of degrees of freedom, for which statistical properties are known exactly.

These systems are defined by a scalar function and two-form, analagous to the construction of Hamiltonian 
systems by a Hamiltonian and a symplectic form.  Indeed, symplectic dynamics arises as a special case of our construction.  
In ergodic cases, the scalar function is interpreted as a probability density, and captures complete information about equal time correlations.  Many dynamical systems share the same probability distribution, but have different un-equal time correlations. Information about the latter also requires the two-form.  We have shown how the Taylor-Mclaurin expansion in time for  un-equal time correlations can be computed without time series simulation.  Replacing a truncated Taylor-Mcalurin expansion with  Pade approximants yields a result which is valid for signicantly greater time seperation.  In principle, it should be possible to determine the large order behavior of the Taylor expansion and infer results about the analytic structure of the correlation functions with respect to time.

A probability density function does not exist for many chaotic systems of interest, namely strange attractors or any dissipative system.  
The dynamical systems which we can construct by the inverse method are both dissipative and anti-dissipative depending on the location within an invariant set.  The dimension is necessarily integer.  We suspect it should be possible to obtain a more general classification, or inverse method, yielding dissipative dynamical systems with fractional dimensions starting with a Hopf characteristic function and a two-form.  The Hopf function exists even in cases where a probability density function does not, in which case the large $J$ asymptotics of the Hopf function are such that its Fourier transform does not converge.

\section*{Acknowledgments}

We wish to thank D. Obeid for discussions. This work was supported in part by funds provided by the US Department of Energy (DOE) grant DE-SC$0010010$ - Task D. C. Pehlevan was supported by a fellowship from the Swartz Foundation.

\end{document}